\documentstyle[12pt]{article}

\rm \textwidth12.6cm
\rm \textheight19.5cm

\begin{document}

\author{{\footnotesize Geng Cheng{\rm \parskip10pt}} \\
{\footnotesize {\it Fundamental Physics Center,USTC., Hefei, Anhui, 230026,
P. R. China}}{\rm \parskip0pt}}
\title{\baselineskip4pt{\small {\bf Advance in dynamical spontaneous symmetry
breaking}}{\rm \parskip0pt}}
\date{}
\maketitle

\begin{abstract}
{\footnotesize {\rm \noindent Recently, a condition is derived for a
nontrivial solution of the Schwinger-Dyson equation to be accompanied by a
Goldstone bound state in a special quantum electrodynamics model. This
result is extended and a new form of the Goldstone theorem is obtained in a
general quantum field theory framework.\parskip4pt }}
\end{abstract}

{\rm \topmargin1pt }

{\rm 
}

{\rm 
}

{\rm {\footnotesize \noindent {\bf I. Introduction} } }

{\rm {\footnotesize Dynamical spontaneous symmetry breaking (DSSB) has been
extremely useful in a wide variety of topics in particle physics,
superconductivity theory (SCT), as well as condensed matter physics.$^{1-3}$ 
} }

{\rm {\footnotesize There are two aspects in the mechanism of DSSB, the mass
generation of fermions and the existence of the Goldstone boson. More
concretely, in particle physics, chiral symmetry breaking occurs when the
self-consistent Schwinger-Dyson (SD) equation for the fermion mass develops
a non-trivial solution as the coupling strength reaches a critical value. In
addition, for the same coupling strength, there must also exist a bound
state solution of the Bethe-Salpeter (BS) equation, with vanishing
4-momenta, corresponding to the massless pseudoscalar Goldstone boson. When
both of these conditions are met, the self-energy solution shall be called a
symmetry-breaking solution. } }

{\rm {\footnotesize Recently, a criterion for a solution of the SD equation
to signal a DSSB is deduced in a special quantum electrodynamics (QED) model.%
$^{1,2}$ In this paper we extend this result to quite a general case. At the
same time a new form of the Goldstone theorem is obtained in a general
quantum field theory framework. } }

{\rm {\footnotesize \parskip10pt\noindent {\bf II. Goldstone theorem. }} }

{\rm {\footnotesize The Goldstone theorem has been established well. We know
from it that a massless boson appears whenever a generator of a global
continuous symmetry group is broken spontaneously. The quantum number of
this particle corresponds to that of the operator of the current which is
broken. For massless quantum field theory, when the mass is generated due to
the self-energy, the chiral symmetry is broken so that the Goldstone
particle is a pseudoscalar boson. We associate this to a bound state
solution of the BS equation, in which the fermion mass will be identified
with the dynamically generated mass from the SD equation. In the following
our discussions will be in this framework.} }

{\rm {\footnotesize The renormalized fermion propagator will be written as $%
S_f^{-1}(p)=\gamma \cdot p~\alpha (p^2)-\beta (p^2).$ The equation satisfied
by the renormalization vertex function is$^4$} }

{\rm {\footnotesize \noindent  
\begin{eqnarray*}
\Gamma _\mu (p^{\prime },p)_{\gamma \delta } &=&Z_1(\gamma _\mu )_{\gamma
\delta }-\int \frac{d^4q}{(2\pi )^4}\left[ S_f(p^{\prime }+q)\Gamma _\mu
(p^{\prime }+q,p+q)S_f(p+q)\right] _{\beta \alpha }\times \\
&&\ \ \times K_{\alpha \beta ,\delta \gamma }(p^{\prime }+q,p+q,q)\qquad \
(1)\ 
\end{eqnarray*}
\noindent Where the $K_{\alpha \beta ,\delta \gamma }(p^{\prime }+q,p+q,q)$
is the fermion-antifermion two body irreducible function. Substitute it in
the renormalized Ward's identity, $(p^{\prime }-p)_\mu \Gamma _\mu
(p^{\prime },p)=S_f^{-1}(p^{\prime })-S_f^{-1}(p)$ and suppose that the $%
K_{\alpha \beta ,\delta \gamma }(p^{\prime }+q,p+q,q)$ dependent on its
argument only through the difference $q=p^{\prime }-p,$ we obtain: } }

{\rm {\footnotesize \noindent  
\[
S_f^{-1}(p)_{\gamma \delta }=Z_1(\gamma _\mu )_{\gamma \delta }+\int \frac{%
d^4p^{\prime }}{(2\pi )^4}S_f(p^{\prime })_{\beta \alpha }\times K_{\alpha
\beta ,\delta \gamma }(p^{\prime }-p)\qquad (2) 
\]
\noindent Taking trace on both sides and similarly with the multiplication
of $\gamma \cdot p$ before trace, the SD equation for the massless fermion
self-energy is simplified into the following set of equations: }}

{\rm {\footnotesize \noindent  
\[
\beta (p^2)=\int \frac{dp^{\prime \,4}}{(2\pi )^4}\frac{\beta (p^{\prime
\,2})}{p^{\prime \,2}\alpha ^2(p^{\prime \,2})+\beta ^2(p^{\prime \,2})}%
\times T_r\left[ K(p^{\prime }-p)\right] \qquad (3) 
\]
}}

{\rm {\footnotesize \noindent  
\[
\alpha (p^2)=1+\frac 1{p^2}\int \frac{dp^{\prime \,4}}{(2\pi )^4}\frac{%
\alpha (p^{\prime \,2})}{p^{\prime \,2}\alpha ^2(p^{\prime \,2})+\beta
^2(p^{\prime \,2})}\left[ (\gamma \cdot p)_{\gamma \delta }(\gamma \cdot
p^{\prime })_{\beta \alpha }K(p^{\prime }-p)_{\alpha \beta ,\delta \gamma
}\right] \quad (4) 
\]
\noindent Here we have used a topological characteristic which leads to a
conclusion that in the $K(p^{\prime }-p)$ the $\gamma _\mu $ appears and
only appears in pair.\ The trivial solution, $\beta (p^2)=0$ and $\alpha
(p^2)$ = finite, corresponds to that of the symmetric vacuum. But it is
certainly true that not all of the nontrivial solutions signal spontaneous
symmetry breaking. The criterion that a nontrivial $\beta (p^2)$ does
trigger DCSB is that there must be an accompanying Goldstone boson. That is
to say, there must be a bound state solution of the BS equation, with the
appropriate quantum numbers. We express the wave function for a bound state
composed of the fermion-antifermion pair of the same kind as $\chi _k(p^2).$
The BS equation is$^5$ }}

{\rm {\footnotesize \noindent  
\[
\left[ S_f^{-1}(\frac k2+p)\chi _k(p^2)S_f^{-1}(\frac k2-p)\right] _{\gamma
\delta }=\int \frac{d^4p^{\prime }}{(2\pi )^4}\chi _k(p^2)_{\beta \alpha
}\times K_{\alpha \beta ,\delta \gamma }(p^{\prime }-p)\qquad (5) 
\]
Since the Goldstone state has $\ell ^P=0^{-}$ and vanishing mass, we take $%
k=0$ and write $\chi _k(p)=\chi _0^P(p^2)\times \gamma _5.$ Now we put
forward an crucial step by requiring the fermion-antifermion two body
irreducible function, $K(p^{\prime }-p),$ to satisfy the relation, $\gamma
_{5\beta \alpha }K(p^{\prime }-p)_{\alpha \beta ,\delta \gamma }=-\left[
K(p^{\prime }-p)\gamma _5\right] _{\gamma \delta }.$ In the same steps as we
do to the SD equation, it leads to the important result: }}

{\rm {\footnotesize \noindent  
\[
\lbrack p^2\alpha ^2(p^2)+\beta ^2(p^2)]\chi _0^P(p^2)=\int \frac{dp^{\prime
\,4}}{(2\pi )^4}\chi _0^P(p^{\prime \,2})\times T_r\left[ K(p^{\prime
}-p)\right] \qquad (6) 
\]
\noindent Introduce the function $\psi (p^2)=[p^2\alpha (p^2)+\beta
^2(p^2)]\chi _0^P(p^2),$ then eq.(6) becomes: }}

{\rm {\footnotesize \noindent  
\[
\psi (p^2)=\int \frac{dp^{\prime \,4}}{(2\pi )^4}\frac{\psi (p^{\prime \,2})%
}{p^{\prime \,2}\alpha ^2(p^{\prime \,2})+\beta ^2(p^{\prime \,2})}\times
T_r\left[ K(p^{\prime }-p)\right] \qquad (7) 
\]
\noindent If there is a solution of eq.(3), then we have a solution of
eq.(7) as $\psi (p^2)=c\beta ^2(p^2).$ c is a arbitrary constant. It give a
solution of the wave function of the Goldstone boson. Thus we obtained a new
form of the Goldstone theorem here. }}

{\rm {\footnotesize \parskip10pt\noindent ${\bf III.}${\bf Condition on the
symmetry breaking solution.} }}

{\rm {\footnotesize For $\chi _0^p$ to be a wave function of a Goldstone
boson which accompanies with the non-trivial solution of the SD equation, it
must satisfy a normal condition:$^5$ }}

{\rm {\footnotesize \noindent  
\[
\int dq^4[\alpha ^2(q^2)q^2+\beta ^2(q^2)]|\chi
_0^P(q^2)|^2=~finite\;and\;nonzero\qquad (8) 
\]
\noindent In order for DSSB to occur, eq.(3) must have a non-trivial
solution $\beta (p^2)\not =0$. However, this is a necessary, but not a
sufficient condition. For $\beta (p^2)$ to be symmetry-breaking solution, it
must satisfy also an additional normalization condition which is obtained
from the normal condition of BS equation. It can be proved that the
necessary and sufficient condition for a non-vanishing solution $\beta (p^2)$
of eq.(3) to be symmetry-breaking is that it must satisfy, together with the
finite solution $\alpha (p^2)$ of eq.(4), the condition: }}

{\rm {\footnotesize \noindent  
\[
\int dq^4\frac 1{q^2\alpha ^2(q^2)+\beta ^2(q^2)}|\beta
(q^2)|^2=~finite\;and\;nonzero.\qquad (9)
\]
}}

{\rm {\footnotesize \parskip6pt\noindent {\bf Acknowledgments} }}

{\rm {\footnotesize G.C. is supported in part by the National Science
Foundation and the NDSTPR Foundation in China. }}

{\rm {\footnotesize \parskip10pt\noindent \baselineskip2pt{\bf References} }}

{\rm {\footnotesize \thinspace \noindent 1. G. Cheng, A Condition on the
Super-Conducting Solution for Schwinger-Dyson Equation in Quenched Planar
QED, In {\it Precision Test of Standard Model and New Physics}, CCAST-WL
Workshop Series {\bf 55,} 73(1996) ed., Chao-hsi Chang. }}

{\rm {\footnotesize \thinspace \noindent 2. G. Cheng and T.K. Kuo, {\it J.
Math. Phys.}{\bf \ 38, }6119(1997). }}

{\rm {\footnotesize \thinspace \noindent 3. Y. Nambu, {\it Phys. Rev.} {\bf %
117}, 648 (1960), G.M. Eliashberg, {\it Soviet Physics JETP}, {\bf 11, }%
686(1960), Y. Nambu and G.J.Lasinio, {\it Phys. Rev.} {\bf 122}, 345 (1961). 
}}

{\rm {\footnotesize \thinspace \noindent 4. J.D. Bjorken \& S.D. Drell, {\it %
Relativistic Quantum Fields, }McGraw-Hill, (1965). }}

{\rm {\footnotesize \thinspace \noindent 5. D. Lurie, {\it Particles and
Fields, }Interscience, (1968). }}

\end{document}